\documentclass{article}
\usepackage[utf8]{inputenc}
\usepackage{amsmath}
\usepackage{amsfonts}
\usepackage{cmbright}
\usepackage{optidef}
\usepackage{bbm}
\usepackage{xcolor}
\usepackage{parskip}
\usepackage{mathtools}
\usepackage{subcaption}
\usepackage{graphicx}
\usepackage[normalem]{ulem}
\usepackage{framed}
\usepackage{comment}
\usepackage{algorithm}
\usepackage{algpseudocode}
\usepackage{authblk}
\usepackage[
doi=false,isbn=false,url=false,eprint=false,
backend=biber,
style=numeric,
sorting=none,
giveninits=true,
maxbibnames=3
]{biblatex}
\renewbibmacro{in:}{}
\AtEveryBibitem{%
  \clearlist{language}%
  \clearfield{pages}
  \clearfield{note}
  \clearfield{month}
}
\title{Dosimetric impact of real-time re-optimization of proton pencil-beam scanning for moving targets}

\author[1,2]{Ivar Bengtsson\thanks{Contact: ivarben@kth.se}}
\author[1]{Anders Forsgren}
\author[2]{Albin Fredriksson}

\affil[1]{Department of Mathematics, KTH Royal Institute of Technology, Stockholm, Sweden}
\affil[2]{RaySearch Laboratories AB, Stockholm, Sweden}

\date{\today}

\begin{document}

\maketitle

\begin{abstract}

    \noindent \textbf{Background:} When treating moving tumors, the precise delivery of proton therapy by \textit{pencil beam scanning} (PBS) is challenged by the \textit{interplay effect}. Although several 4D-optimization methods have been implemented in the literature and commercial systems, what is the most beneficial motion management technique is still an open question.
\\\\
   \textbf{Purpose:} In this study, we wish to investigate the dosimetric impact of re-optimizing the PBS spot weights at multiple instances during the treatment delivery in response to, and in anticipation of, variations in the patient's breathing pattern.
\\\\
   \textbf{Methods:} We simulate for PBS the implementation of a real-time adaptive framework based on principles from receding horizon control. We consider the patient motion as characterized by a one-dimensional amplitude signal and a 4DCT, to simulate patient breathing of variable frequency. The framework tracks the signal and predicts the future motion with uncertainty increasing with the length of the prediction horizon. After each delivered energy layer, the framework re-optimizes the spot weights of the next layer based on the delivered dose and the predicted motion. For three lung patients, we generate 500 variable breathing patterns to evaluate the dosimetric results of the framework and compare them to those of implementations of previously proposed non-adaptive methods.
\\\\
    \textbf{Results:} Compared to the best non-adaptive method, the adaptive framework improves the CTV D98 in the near-worst breathing scenario (5th percentile), from  96.4 to 98.9 \% of the prescribed dose and considerably reduces the variation as measured by a decrease in the inter-quartile range by more than 80 \%, when averaged over the patients. The target coverage improvements are achieved without generally compromising target dose homogeneity or OAR dose.
\\\\
    \textbf{Conclusions:} The study indicates that a motion-adaptive approach based on re-optimization of spot weights during delivery has the potential to substantially improve the dosimetric performance of PBS given fast and accurate models of patient motion.

\end{abstract}

\section{Introduction}

In proton treatments, \textit{pencil beam scanning} (PBS) is the preferred delivery technique due to its ability to shape dose distributions with sharp gradients through intensity-modulation \cite{lomax_treatment_2004}. As the accurate delivery of PBS is challenged by patient density and positioning uncertainties, robust optimization is commonly used \cite{fredriksson_minimax_2011}. For tumor sites such as lung and liver, patient motion is yet another concern. \textit{4D-robust optimization} (4DRO), based on image information from a 4DCT, has been proposed to address this problem, although a recent review concluded that its clinical advantage remains to be shown, given the resulting increase in computation time \cite{knopf_clinical_2022}. Furthermore, it does not explicitly account for what is known as the \textit{interplay effect}: the interference between the time structure of the delivery and that of the patient motion \cite{lambert_intrafractional_2005, seco_breathing_2009, bert_quantification_2008, bert_motion_2011}.

Several 4D-optimization methods explicitly considering interplay using \textit{4D dose calculation} (4DDC) have also been proposed. Although these methods have sometimes been called \textit{dynamic} 4D optimization, we will hereafter refer to them as \textit{interplay-driven optimization} (IPO), similar to the terminology introduced by Engwall et al.\ \cite{engwall_4d_2018}. It was shown by Bernatowicz et al.\ that IPO can produce plans with dose distributions of comparable quality (evaluated with motion) as those of 3D-optimized plans (evaluated without motion) if the motion is known a priori, but also that the resulting plans are susceptible to deviations from the assumed motion pattern \cite{bernatowicz_advanced_2017}. To address this, Engwall et al.\ proposed \textit{interplay-robust optimization} (IPRO), which explicitly incorporates a multitude of breathing motion scenarios in the optimization \cite{engwall_4d_2018}. However, IPRO is computationally demanding, given the many scenarios needed to achieve robust plans.

Beyond addressing motion uncertainty and interplay at the treatment planning stage, there have been efforts to enable the adaptation of the treatment delivery in real time. For example, beam tracking is an approach that has reached clinical applications in photon therapy but not yet in proton therapy despite considerable research interest \cite{bert_dosimetric_2010, zhang_online_2014, riboldi_real-time_2012}. Moreover, the group at GSI \textit{Helmholtz Centre for Heavy Ion Research} have proposed strategies where the \textit{treatment control system} (TCS) actively controls which spots are delivered during which phase. One is \textit{multigating}, which, similarly to IPO, optimizes the spot weights using 4DDC but provides a solution to the sensitivity to motion uncertainty by maintaining the spot-to-phase assignment during delivery by gating the beam when necessary \cite{graeff_multigating_2014}. A disadvantage of multigating is that failure to efficiently synchronize the beam with the motion may severely increase treatment times. Another strategy, which was implemented experimentally for rotating motion, is the dose-compensation functionality that modifies future spot weights based on a look-up table \cite{luchtenborg_experimental_2011}. A third is to create a plan library with a uniform dose plan for each motion phase \cite{graeff_motion_2014}. The TCS then uses information about the current phase to deliver parts of the plans in alternation. This approach was implemented experimentally in the dose delivery system at \textit{Centro Nazionale di Adroterapia Oncologica} (CNAO) in Lis et al.\ \cite{lis_modular_2020}. The approach has since then been extended with beam tracking of the tumor to compensate for residual motion not included in the plan library \cite{steinsberger_experimental_2023}. The use of uniform dose plans implies that the method does not need to rely on deformable image registration (DIR), removing an uncertainty source that is otherwise often present in 4D-optimization approaches. However, the required uniformity of the phase doses also results in the loss of some of the degrees of freedom associated with intensity modulation.

In this work, we consider the dosimetric potential of a real-time re-optimization approach for modifying the remaining spot weights during PBS delivery. Intuitively, such an approach would implicitly achieve tracking by changing the spot weights depending on which spots still result in desirable dose distributions given the delivered dose and the estimated and predicted anatomical information. Such a framework has previously been proposed for tomotherapy under the name \textit{motion-adaptive optimization} \cite{lu_real-time_2009}. In this work, we present and simulate the implementation of a similar framework -- derived from \textit{receding horizon control} -- for proton PBS. The framework involves the TCS by modifying the planned spot weights during the treatment delivery. During the delivery, new weights for the spots not yet delivered are determined from solutions to optimization problems that vary based on the observed and anticipated motion. The receding horizon refers to considering for re-optimization only a limited subset of the spots not yet delivered, to limit computation time. The dose delivered by the spots in the horizon is then summed with estimations of the delivered dose and the future dose from spots outside the planning horizon. To our knowledge, our study is the first to investigate online re-optimization of spot weights for PBS. To analyze the framework's dosimetric impact, we performed the simulations for various breathing scenarios for three non-small cell lung cancer patients and compared the results against those from IPO and IPRO. Although much effort remains to make the software and hardware developments needed for a clinically feasible real-time implementation of the approach, we believe that investigating its dosimetric implications is necessary to justify such efforts.

\section{Method}

We first describe the patient motion models and delivery time structure used in this work. Second, we describe the re-optimization framework and the numerical experiments used to evaluate it.

\subsection{Patient motion model}

We assume that the patient's breathing induces the motion, with states corresponding to the phases of a pre-treatment 4DCT. The state $s_t$ at a time $t$ is then determined by amplitude binning of a one-dimensional signal $a(t)$. For a breathing cycle starting at time 0, we use the model from Lujan et al.\ \cite{lujan_method_2003}:

\begin{equation}
    a(t) = A \sin^{2n} (\frac{\pi t}{\tau}), \quad t \in [ 0, \tau ).
\end{equation}

\begin{figure}
    \centering
    \includegraphics[width=\textwidth]{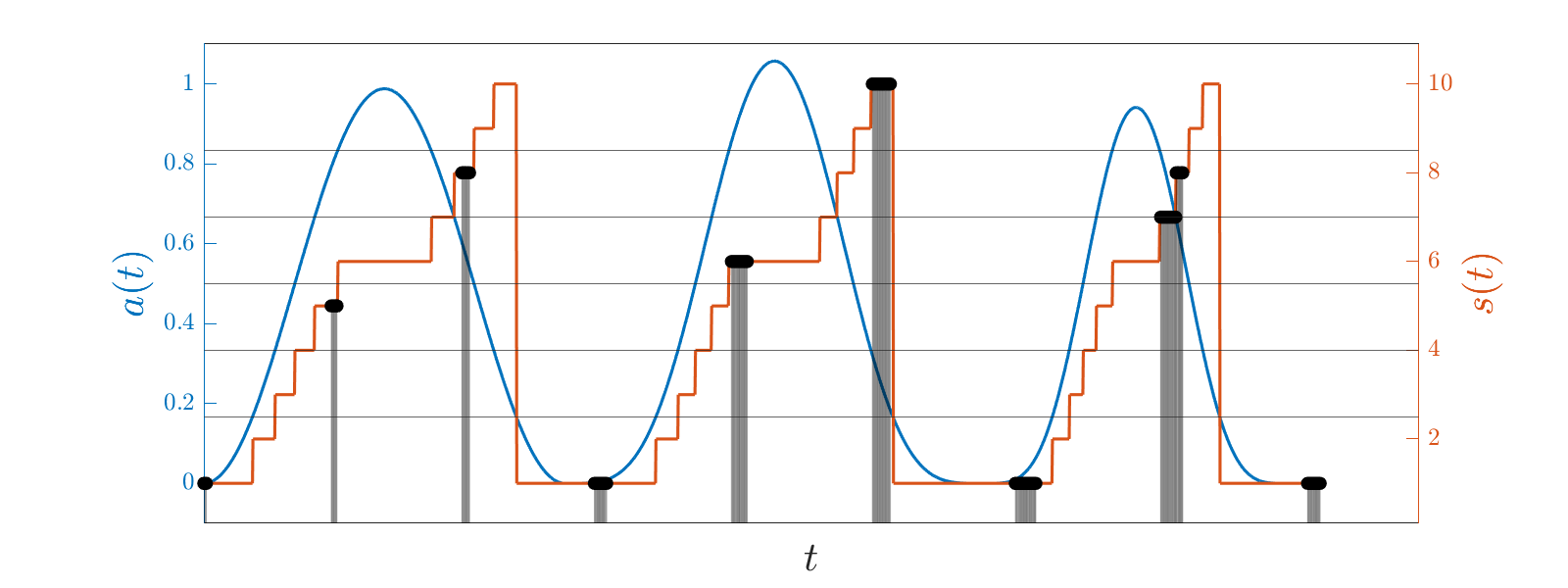}
    \caption{A representation of three consecutive breathing cycles sampled from the motion model and the resulting spot-to-phase assignment. The horizontal lines illustrate the bins used to convert the continuous signal $a(t)$ (blue) into the discrete state $s(t)$ (orange). The horizontal placement of each vertical bar (gray) represents the delivery time of a single spot, and nearby bars constitute an energy layer. The height and the marker (black) at the top of each bar coincide with the state during which the spot is delivered.}
    \label{example_breathing}
\end{figure}

Here, $A$, $n$, and $\tau$ are random variables denoting the maximal amplitude, the degree of asymmetry, and the breathing period, respectively. Treatment breathing signals are then simulated by concatenating signals of independently sampled breathing cycles until the total duration of the signal exceeds the treatment time (Figure \ref{example_breathing}). We assume that each treatment beam can always be started in sync with the breathing cycle.

\subsection{Dose delivery model}

The 4DDCs are based on phase sorting, which works by accumulating the phase doses to a reference phase by deformable registration \cite{bert_4d_2007, engwall_4d_2018}. Each phase dose is computed considering only the spots that are delivered during the occurrences of that phase during the treatment duration. In our model, the dose from a pencil beam spot is dependent only on the spot weight, measured in \textit{monitor units} (MUs) and the motion phase at the spot delivery's start time. This approximation disregards the duration of the spot delivery but should be rather mild as the temporal resolution of a 4DCT implies that the duration of a phase is much greater than that of a single spot and, very few, if any, spot durations will overlap with more than one phase. Thus, a vector $p_k^{s_{t_k}}$, in which the $i$:th element denotes the dose per MU to voxel $i$, is used to model the dose deposited by spot $k$ when delivered at time $t_k$ on phase $s_{t_k}$ and then deformed to the reference phase. We then model the process of depositing dose by PBS as a system with discrete time steps corresponding to individual pencil beam spots:

\begin{equation}
    d_k = d_{k-1} + p_k^{s_{t_k}} x_k.
\end{equation}

The delivery time structure model used replicates the one from Pfeiler et al.\ \cite{pfeiler_experimental_2018}. Besides the spot weight $x_k$ (MU), the time $t_{k+1} - t_k$ depends on three parameters: the dose rate $\alpha(x_k)$ (ms/MU); the scanning time $\beta_k$ (ms) for moving the pencil beam between the lateral positions of spot $k$ and spot $k+1$; and the energy switching time $\gamma_k$ (ms), which is non-zero if and only if spot $k$ is the last within its energy layer (Figure  \ref{example_breathing}).

\subsection{Interplay-adaptive optimization}

In light of the introduced models, we view PBS planning with IPO, after already having delivered $\hat{k}$ spots, as solving the discrete-time optimal control problem:

\begin{mini}|l|
{x \in \mathbb{R}^n}{f(d_n)}{}{}
\addConstraint{d_k}{= d_{k-1} + p_k^{s_{t_k}} x_k}{k = \hat{k}+1, \dots, n}
\addConstraint{t_k}{= t_{k-1} + \alpha(x_{k-1}) x_{k-1} + \beta_{k-1} + \gamma_{k-1}, \quad}{k = \hat{k}+2, \dots, n}
\addConstraint{x_k}{\in \{ 0 \} \cup [ l, \infty ), \quad}{k = \hat{k}+1, \dots, n},
\label{pbs_control_problem}
\end{mini}

where the delivered dose $d_{\hat{k}}$, the start time of the upcoming spot $t_{\hat{k}+1}$, and the spot weight lower bound $l$ are given. The uncertainty in Problem \ref{pbs_control_problem} resides in each $s_{t_k}$, which depends on the random variables from the breathing model, and increases with $t_k$, which leads to increased uncertainty when planning far ahead \cite{bernatowicz_advanced_2017}. An intuitive remedy to this uncertainty would be to continuously track and predict patient motion and repeatedly solve instances of Problem \ref{pbs_control_problem} during delivery. This adaptive approach would allow soon-to-be-delivered spots to be re-optimized with less uncertainty about their phase assignment.  The size of Problem \ref{pbs_control_problem}, however, is comparable to those of ordinary treatment planning optimization problems and results in computation times that inhibit a real-time implementation on current hardware. We therefore use an RHC-based approach with horizon length $h_{\hat{k}}$, that satisfies $h_{\hat{k}} << n$, to reduce the number of considered spots
and consequently the number of optimization variables. Since the time dependency on $x$ is small compared to the time per phase in the 4DCT, we treat $t$ as constant by fixing $t(x) = t^0 = t(x^0)$, where $x^0$ is the initial solution. Preferably, $x^0$ corresponds to a deliverable and near-optimal pre-computed plan. Beyond the optimization horizon, the \textit{predicted remaining dose}, $d_{\hat{k}}^{\text{rest}}$, is considered as a constant during the optimization and defined as the dose that would be delivered if the predicted motion pattern was correct and no further modifications were made to the spot weights:

\begin{equation}
    d_{\hat{k}}^{\text{rest}} \coloneqq \sum^n_{k = \hat{k} + h_{\hat{k}} + 1} p_k^{s_{t^0_k}} \cdot x_k^0.
\end{equation}

Finally, the lower bound on $x$ is enforced for all elements:

\begin{mini}|l|
{(x_{\hat{k}+1}, \dots, x_{\hat{k}+h_{\hat{k}}})}{f \left(d_{\hat{k}} + \sum_{k=\hat{k}+1}^{\hat{k}+h_{\hat{k}}} p_k^{s_{t^0_k}}x_k \quad + d_{\hat{k}}^{\text{rest}} \right)}{}{}
\addConstraint{x_k}{\geq l, \quad}{k = \hat{k}+1, \dots, \hat{k}+h_{\hat{k}}}.
\label{pbs_mpc_problem}
\end{mini}

For an ordered set of spot indices $K$ after which to solve Problem \ref{pbs_mpc_problem}, and the corresponding set of horizon lengths $H \coloneqq \{ h_{\hat{k}} \}_{\hat{k} \in K}$, the framework follows Algorithm \ref{alg}. The implementation details are described in Section \ref{adaptive_details}.

\begin{algorithm}
\caption{Interplay-adaptive optimization (IPAO)}
\begin{algorithmic}[1]
  \State Initialize current spot weights $\Tilde{x}$
        \For{$\hat{k} \in K$}
        \State Set initial point: $x^0 \leftarrow \Tilde{x}$.
          \State Compute spot start times: $\{ t^0_1, \dots, t^0_n \} \leftarrow t(x^0)$.
          \State Predict the future motion $\{ s_{t^0_{\hat{k}+1}}, \dots, s_{t^0_n} \}$.
          
          \State Compute delivered dose: $d_{\hat{k}} \leftarrow d( x^0_1, \dots, x^0_{\hat{k}}, s_{t^0_1}, \dots, s_{t^0_{\hat{k}}})$.
          \State Compute remaining dose: $d_{\text{rest}} \leftarrow d(x^0_{\hat{k}+\hat{h}+1}, \dots, x^0_{n}, s_{t^0_{\hat{k}+\hat{h}+1}}, \dots, s_{t^0_n}) $.
            \State $x^{\star} \leftarrow$ Solve Problem \ref{pbs_mpc_problem} with initial point $x^0$.
            \State Set current spot weights: $\Tilde{x}_k \leftarrow x^{\star}_k, \quad k = \hat{k}+1, \dots, \hat{k}+h_{\hat{k}} $.
  \EndFor
  
\end{algorithmic}
\label{alg}
\end{algorithm}

\subsection{Patient data}

Three 10-phase 4DCTs, from a data set of lung patients uploaded to The Cancer Imaging Archive (TCIA) \cite{clark_cancer_2013}, were used in the numerical experiments. The complete data set is described in detail in Hugo et al.\ \cite{hugo_longitudinal_2017}. Breathing regularity was upheld with the help of audio-visual feedback, and patient-specific motion magnitudes and breathing cycle statistics are presented in Table \ref{patients}. The patients used in this study were selected to consider different motion magnitudes and CTV volumes.

\begin{table}[h!]
    \centering
    \begin{tabular}{c|c|c|c|c|c}
    \hline
        ID & Motion (cm) & $\mu_{\tau}$ (s) & $\sigma_{\tau}$ (s) & CTV volume ($\text{cm}^3$) & No. spots \\ \hline
        P101 & 0.74 & 3.70 & 0.42 & 29.93 & 2708 \\ \hline
        P111 & 1.22 & 3.20 & 0.16 & 71.52 & 3622 \\ \hline
        P114 & 0.97 & 3.20 & 0.26 & 185.30 & 9072 \\ \hline
    \end{tabular}
    \caption{Characteristics of each patient. Motion is measured as the mean displacement vector length within the ITV when registering the maximum expiration phase to the maximum inspiration phase. The mean and standard deviation of the breathing period $\tau$ were computed from the breathing periods during image acquisition. The CTV volumes were computed on the end-inhalation phase.}
    \label{patients}
\end{table}

We used the RayStation (RaySearch Laboratories AB, Stockholm, Sweden) Monte Carlo dose engine (version 5.3) to generate pencil beam spot data for each phase \cite{janson_treatment_2024}. Deformable image registration was then performed with RayStation's algorithm ANACONDA between a reference phase and the other phases in the 4DCT \cite{weistrand_anaconda_2015}. The registrations were then used to map the dose deposition vectors on each phase to the voxel grid of the reference phase.

\subsection{Numerical experiments}
Breathing signals $\{ a(t) \}$ were generated by sampling each of the parameters $A, n,$ and $\tau$ for multiple breathing cycles, which were then concatenated. $A$ was taken from a truncated normal distribution with $\mu_A = 1, \sigma_A = 0.05$. Likewise, $\tau$ was taken from a truncated normal distribution but with patient-specific mean and standard deviation (Table \ref{patients}). Both distributions were truncated to within two standard deviations from the mean. Finally, $n$ could take any value in $\{ 1, 2, 3 \}$ with equal probability.

All delivery time structure model parameters were taken from Pfeiler et al.\ \cite{pfeiler_experimental_2018}. Their dose rate model for $\alpha(\cdot)$ (ms/MU) was replicated exactly as: $\alpha(x) = \frac{2588.986}{(x+1.256)^{16.721}} + 4.995$. The scanning times in the lateral coordinates $(y, z)$ were similarly taken as $\beta^{y}(\Delta y) = 0.3125 \Delta y + 2.2187$ and $\beta^{z}(\Delta z) = 3.1250 \Delta z + 1.9375$ (ms/cm), respectively. The inter-spot scanning time was then $\beta_k = \text{max}\{ \beta^y(|y_k - y_{k+1}|), \beta^z(|z_k - z_{k+1}|) \}$. The energy switching time was treated as a constant, $\gamma = 1230$ ms.

To treat the motion mitigation in isolation, no setup or range uncertainties were considered. All optimization problems considered the same objectives: a minimum-dose objective penalizing dose below the prescription (200 cGy / fx) in the CTV; a maximum-dose objective penalizing dose more than 5\% above the prescription in a 1 cm expansion of the CTV; and three low-weighted maximum-dose objectives penalizing dose to the heart, healthy lung (right lung - CTV), and esophagus, respectively (Table \ref{objectives}). Across all experiments, the optimizations were warm-started from the spot-weight vector of a conventional 4DRO plan, optimized with 40 iterations in RayStation \cite{janson_treatment_2024}, for which spot filtering had been applied during optimization to impose a minimum spot-weight bound of $l = 0.02$ monitor units. We then enforced this bound in all optimizations to ensure that all plans were deliverable.

\begin{table}[h!]
    \centering
    \begin{tabular}{c|c|c|c|c}
    \hline
        ID & ROI & Type & Fx.\ dose level (cGy) & Weight \\ \hline
        1 & CTV & Minimum dose & 200 & 100  \\ \hline
        2 & CTV+1cm & Maximum dose & 210 & 50  \\ \hline
        3 & Heart & Maximum dose & 66.67 & 1  \\ \hline
        4 & Esophagus & Maximum dose & 100 & 1  \\ \hline
        5 & Healthy lung & Maximum dose & 166.67 & 1  \\ \hline
    \end{tabular}
    \caption{The parameters of the optimization functions. The definitions of the objective types mimic those in RayStation, with squared penalties of dose deviations summed over all voxels in an ROI.}
    \label{objectives}
\end{table}

Except for optimizing the 4DRO plan, all numerical experiments were performed in Matlab (2024a, Mathworks Inc, Natick, Massachusetts, USA). Specifically, optimization problems were solved using an API to the \textit{sequential quadratic programming} (SQP) solver SNOPT (7.7 Stanford Business Software, Stanford, California).

\subsubsection{Adaptive approach (IPAO)} \label{adaptive_details}

IPAO was implemented as described in Algorithm \ref{alg}, with $K$ and $H$ chosen to re-optimize the spot weights on a per-energy-layer basis. More precisely, $K$ consisted of the indices of the last spot in each energy layer, while the corresponding horizon lengths in $H$ were the number of spots in the subsequent energy layer. The main principle behind this hyperparameter choice was that the beam-off period during an energy layer switch is a natural time to perform computations within a potential real-time implementation of the framework. Each optimization problem was solved using 20 SQP iterations, after which a notable decrease in the objective function was no longer observed.

\begin{figure}
    \centering
    \begin{subfigure}{0.49\textwidth}
        \centering
        \includegraphics[width=\linewidth]{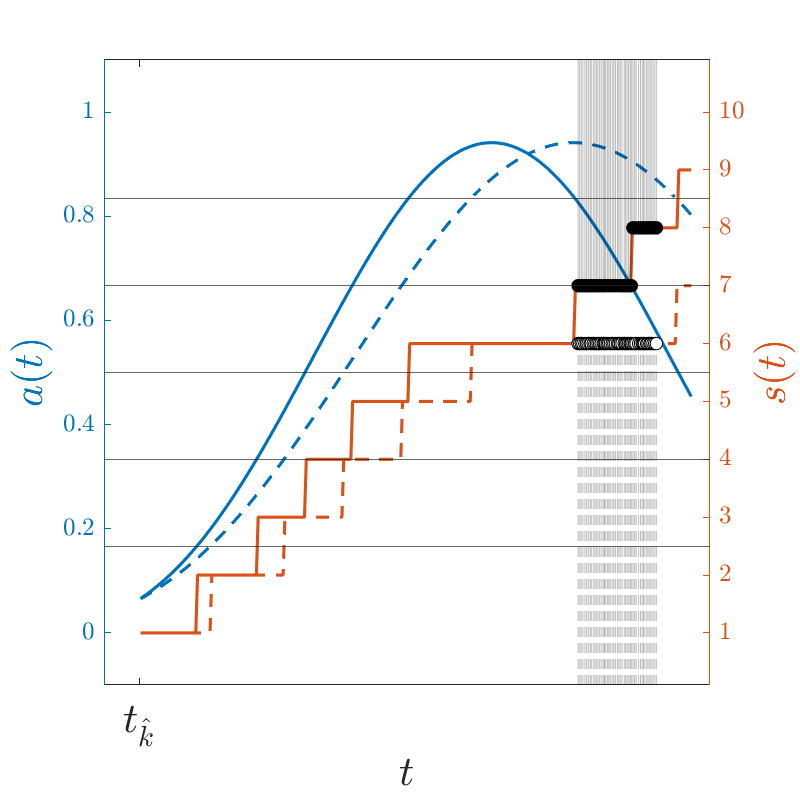}
        \caption{An incorrect prediction of the spot-to-phase assignment in the energy layer following spot $\hat{k}$.}
        \label{fig:sub1}
    \end{subfigure}
    \hfill
    \begin{subfigure}{0.49\textwidth}
        \centering
        \includegraphics[width=\linewidth]{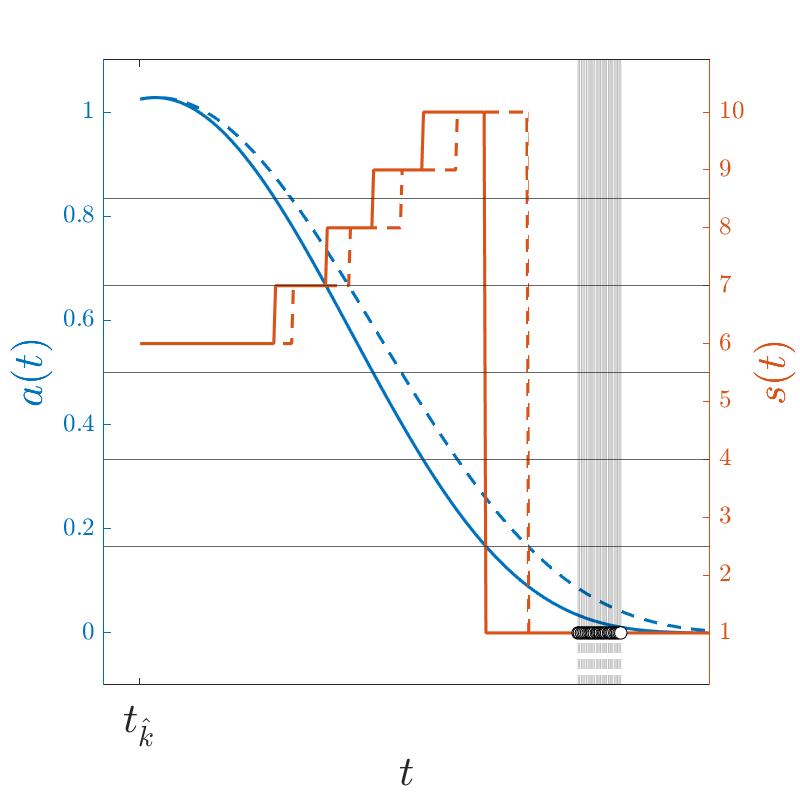}
        \caption{A correct prediction of the spot-to-phase assignment in the energy layer following spot $\hat{k}$.}
        \label{fig:sub2}
    \end{subfigure}
    \caption{Examples of the motion prediction model in the adaptive approach. The solid lines represent the actual motion signal $a(t)$ (blue) and the actual discrete state $s(t)$ (orange). The dashed lines represent the corresponding predictions. The horizontal placements of the vertical bars indicate the start times of the spots in the next energy layer, and the markers at the end of each line coincide with the state at the spot's delivery time.}
    \label{example_prediction}
\end{figure}

We evaluated IPAO 500 times, each considering a different simulation of the breathing signal $a(t)$. The number of evaluation scenarios was chosen as suggested by Pastor-Serrano et al.\ to obtain sufficient statistical accuracy for the evaluation of interplay effects \cite{pastor-serrano_how_2021}. At each optimization, Algorithm \ref{alg} had access to the past motion and made a simple prediction of the future. Predictions of the future motion assumed exact knowledge of the current values of $a, A,$ and $n$, while the period $\tau$ was assumed to take its patient-specific mean value (Table \ref{patients}). Naturally, the benefit of adaptation depends on the accuracy of these predictions. In our experiments, the assumptions above resulted in the prediction being correct for approximately 70\% of the spots. Figure \ref{example_prediction} illustrates how the predictions were made.

\subsubsection{Non-adaptive approaches (IPO and IPRO)}

To establish benchmark results with which to compare the adaptive framework, non-adaptive IP(R)O optimization approaches, as proposed in Bernatowicz et al.\ and Engwall et al., were implemented as special cases of Algorithm \ref{alg}, with $K = \{0\}$ and $H = n$ \cite{bernatowicz_advanced_2017, engwall_4d_2018}. In the robust case, the objective function was computed by taking the maximal (smoothly approximated by the weighted power mean with parameter 8) sum of the robust CTV objectives over the nominal and 39 other independently sampled motion scenarios and then adding the nominal values of the non-robust objectives. As the solution relies on the assumption of fixed time structures, the iterative scheme described in Algorithm \ref{alg2} was used to update the spot start times at fixed intervals. Each IP(R)O method was optimized with $N = 20$ and 10 iterations per instance of Problem \ref{pbs_mpc_problem} (resulting in a total of 200 SQP iterations). For reference, the 4DRO plan from RayStation was also included in the comparison. Each of the non-adaptive plans were then evaluated on the same 500 breathing scenarios used to evaluate IPAO.

\begin{algorithm}
\caption{Iterative interplay-driven optimization}
\begin{algorithmic}[1]
  \State Input: Initial spot weights $x^0$. Number of updates $N$.
  \State Output: Final spot weights $x^{\star}$.
      \For{$i = 1, \dots, N$}
        \State Compute spot start times $t(x^{i-1})$.
        \State $x^i \leftarrow$ Solve Problem \ref{pbs_mpc_problem} with initial point $x^{i-1}$ and spot start times $t(x^{i-1})$.
      \EndFor
      \State $x^{\star} \leftarrow x^N$.
\end{algorithmic}
\label{alg2}
\end{algorithm}

\section{Results}

\begin{figure}
    \centering
    \begin{subfigure}{0.32\textwidth} 
        \centering
        \includegraphics[width=\linewidth]{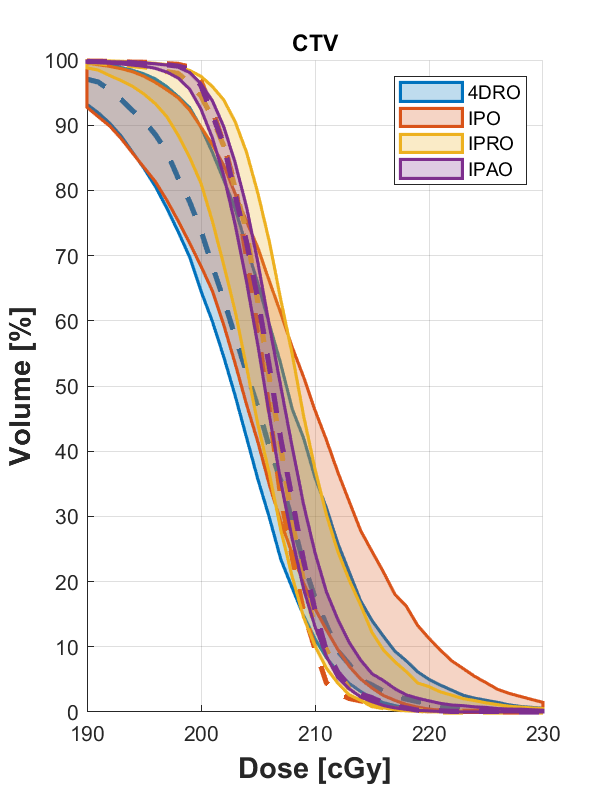}
        \caption{P101}
        \label{CTV_D98_P101}
    \end{subfigure}
    \begin{subfigure}{0.32\textwidth}
        \centering
        \includegraphics[width=\linewidth]{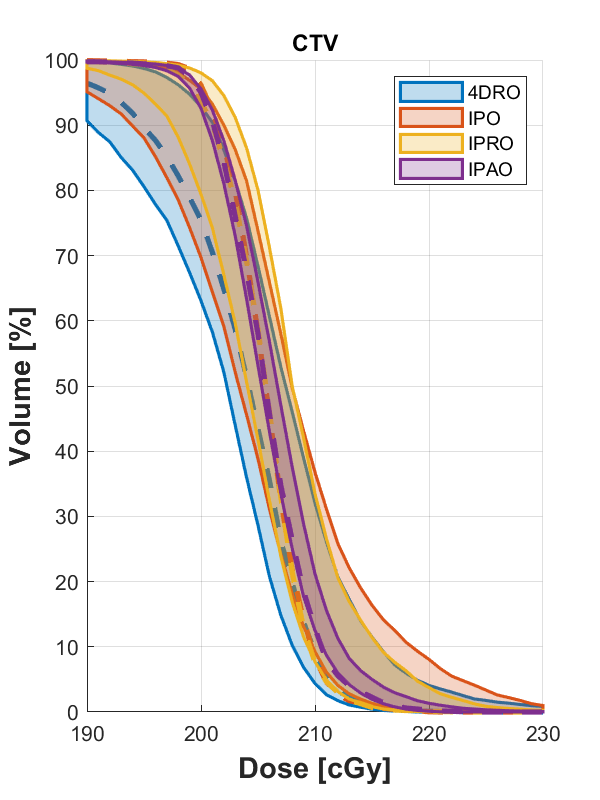}
        \caption{P111}
        \label{CTV_D98_P111}
    \end{subfigure}
    \begin{subfigure}{0.32\textwidth}
        \centering
        \includegraphics[width=\linewidth]{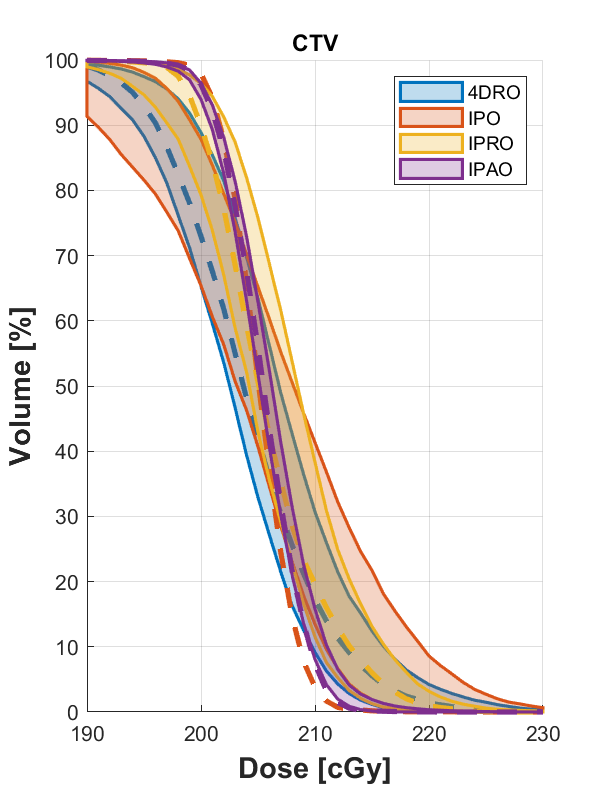}
        \caption{P114}
        \label{CTV_D98_P114}
    \end{subfigure}
    \caption{CTV DVH bands for each of the evaluated methods. At each dose value $\hat{d}$, a DVH band represents a 95\% confidence interval for the value of $V_{\hat{d}}$. For each method, the dashed line represents the DVH in the nominal motion scenario.}
    \label{dvh_bands}
\end{figure}

\begin{figure}
    \centering
    \begin{subfigure}{0.32\textwidth} 
        \centering
        \includegraphics[width=\linewidth]{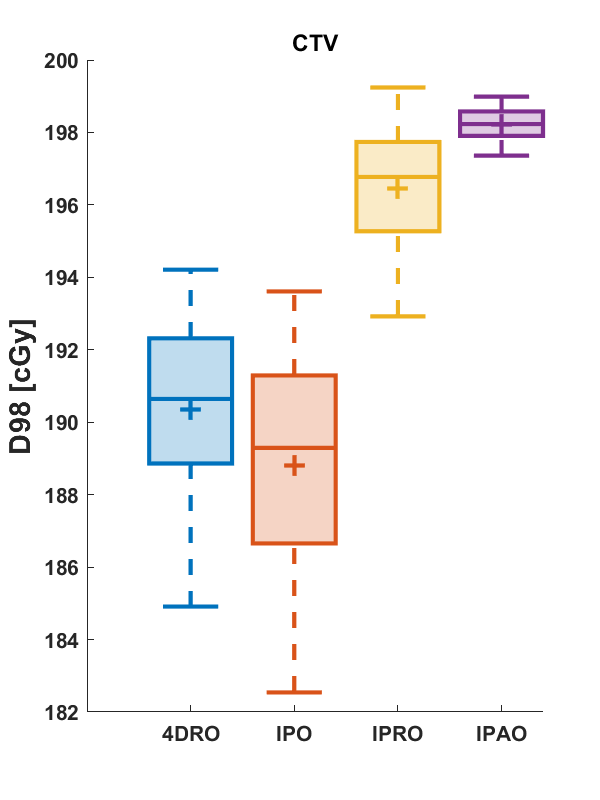}
        \caption{P101}
        \label{P101_ctv_d98}
    \end{subfigure}
    \begin{subfigure}{0.32\textwidth}
        \centering
        \includegraphics[width=\linewidth]{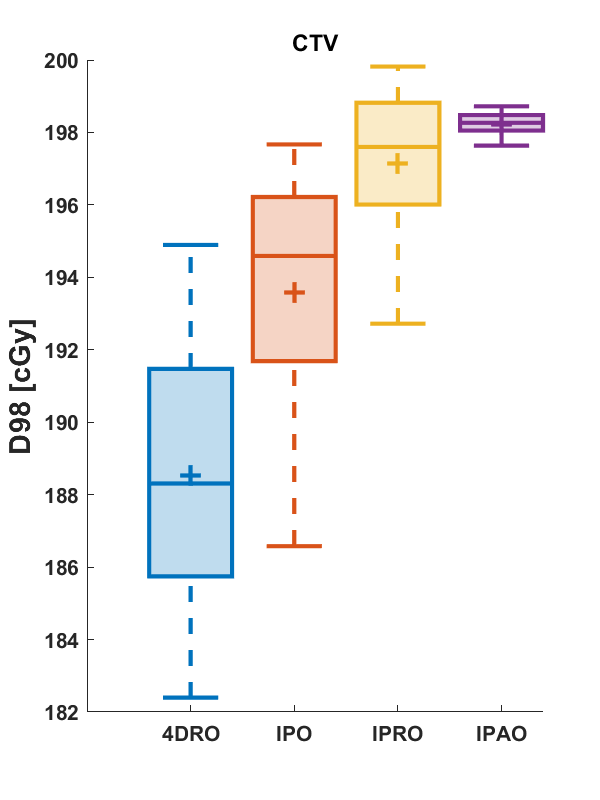}
        \caption{P111}
        \label{P111_ctv_d98}
    \end{subfigure}
    \begin{subfigure}{0.32\textwidth}
        \centering
        \includegraphics[width=\linewidth]{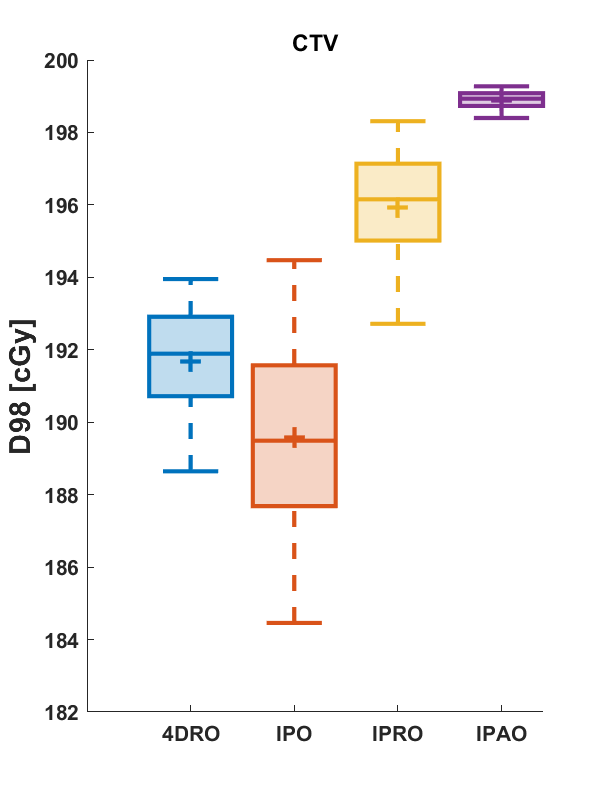}
        \caption{P114}
        \label{P114_ctv_d98}
    \end{subfigure}
    \hfill
    \begin{subfigure}{0.32\textwidth} 
        \centering
        \includegraphics[width=\linewidth]{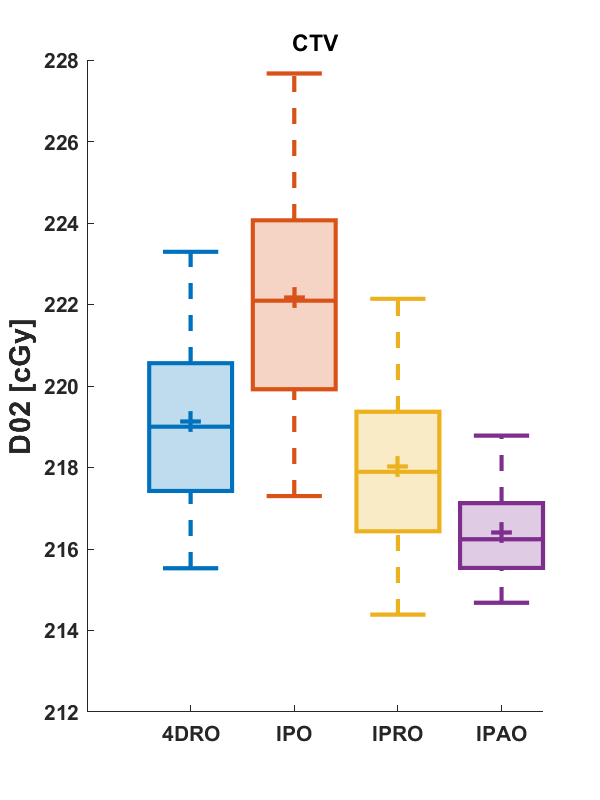}
        \caption{P101}
        \label{P101_ctv_d2}
    \end{subfigure}
    \begin{subfigure}{0.32\textwidth}
        \centering
        \includegraphics[width=\linewidth]{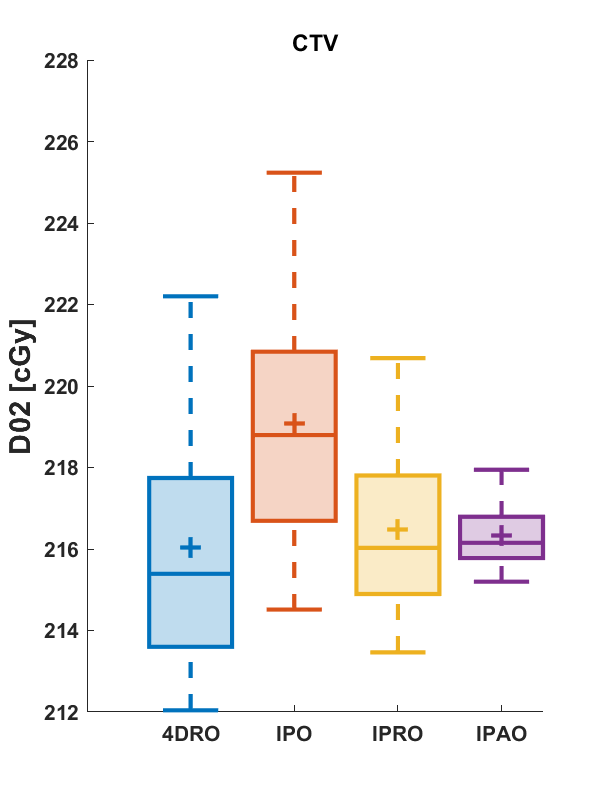}
        \caption{P111}
        \label{P111_ctv_d2}
    \end{subfigure}
    \begin{subfigure}{0.32\textwidth}
        \centering
        \includegraphics[width=\linewidth]{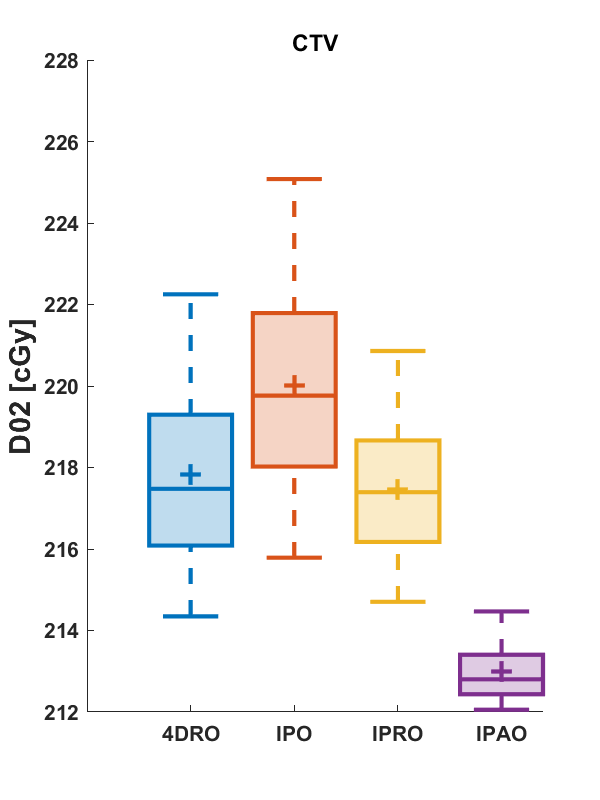}
        \caption{P114}
        \label{P114_ctv_d2}
    \end{subfigure}
    \caption{Box plots showcasing the spread of the CTV D98 and the CTV D2. The box edges indicate quartiles, while the whiskers indicate 5th and 95th percentiles. The mean and median are indicated by the plus and the solid line, respectively.}
    \label{boxplots_ctv}
\end{figure}

Compared to the non-adaptive methods, the CTV DVH band of IPAO is notably steep and narrow. (Figure \ref{dvh_bands}). Across the three patients, the \textit{near-worst-case} CTV D98s (5th percentile, indicated by the lower whiskers in Figures \ref{P101_ctv_d98} - \ref{P114_ctv_d98}) were between 98.68 and 99.20 \% of the prescribed dose (200 cGy / fx). In comparison, the corresponding values were in the intervals 96.36--96.46 and 91.20--94.32 for IPRO and 4DRO, respectively. Another apparent feature of IPAO was the reduction in variation, which was only 15.23--27.36 \% of that of IPRO, when measured by the inter-quartile range. Additionally, the mean CTV D2 was typically kept low and outperformed the other methods in mean for P101 and P114. For P111, the mean CTV D2 was 0.2 \% worse than the lowest among the other methods (4DRO). Like for D98, the reduced variability was apparent for IPAO, as the inter-quartile range was consistently reduced compared to any of the non-adaptive methods. Finally, the nominal DVH of IPAO for the CTV was notably similar to that of IPO. In particular, the relative decrease was between 0.18 and 0.35 \% for D98, and the relative increase in D2 was between 0.38 and 1.30 \%. Minor differences can be spotted in the near-max doses of the DVHs (Figure \ref{dvh_bands}).

A comparison between the non-adaptive methods is also of interest. IPO resulted in a very steep and favorable DVH curve in the nominal scenario (Figure \ref{dvh_bands}). However, the corresponding DVH band indicates substantially worse performance under variation of the patient motion. Compared to 4DRO, which does not explicitly consider any patient motion pattern, the spread of the CTV D98 was similar, while the D2 was typically higher (Figure \ref{boxplots_ctv}). Understandably, the patient for which IPO achieved better target coverage than 4DRO was P111, for which the variability of the breathing period was the lowest (Table \ref{patients}). Compared to IPRO, however, the CTV doses were consistently worse for IPO in terms of both mean and near-worst case (5th and 95th percentiles) for D98 as well as D2. These simultaneous advantages of IPRO show that robustness to target coverage is achieved without compromising target dose homogeneity. For CTV D98, the inter-quartile ranges of IPO were worse than those of IPRO by factors between 1.61 and 1.88.

\begin{figure}
    \centering
    \begin{subfigure}{0.32\textwidth} 
        \centering
        \includegraphics[width=\linewidth]{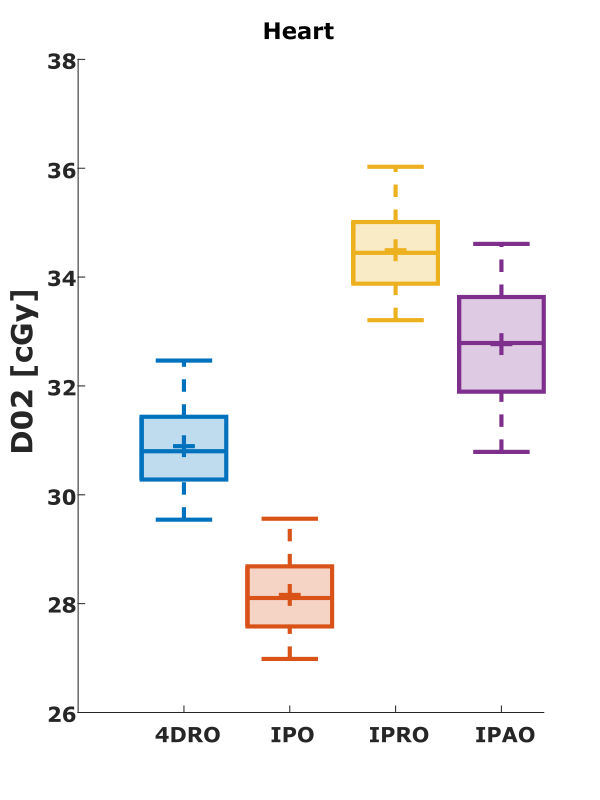}
        \caption{P101 - Heart}
        \label{P101_heart_d2}
    \end{subfigure}
    \begin{subfigure}{0.32\textwidth}
        \centering
        \includegraphics[width=\linewidth]{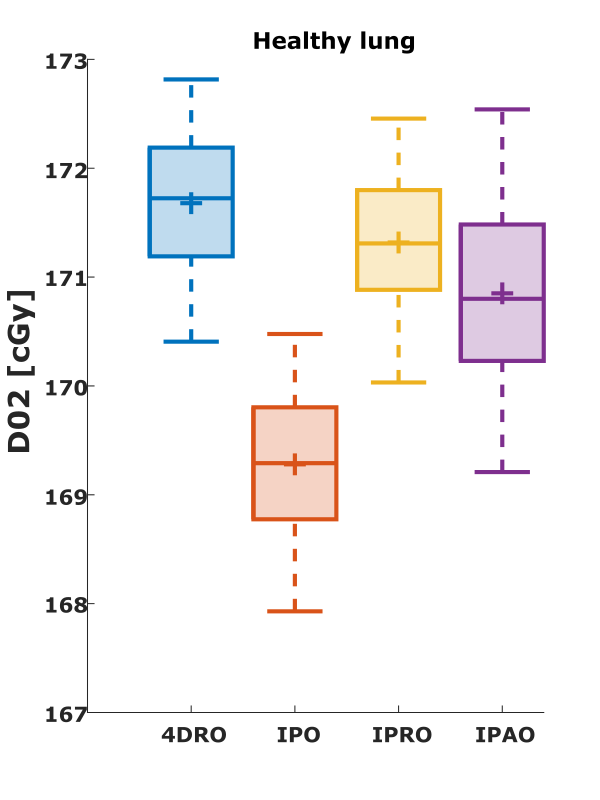}
        \caption{P101 - Healthy lung}
        \label{P101_lung_d2}
    \end{subfigure}
    \begin{subfigure}{0.32\textwidth}
        \centering
        \includegraphics[width=\linewidth]{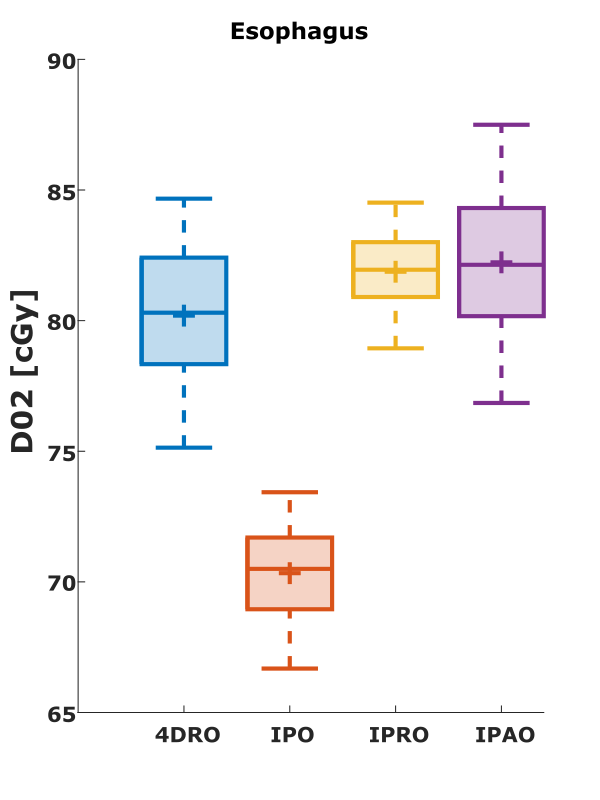}
        \caption{P101 - Esophagus}
        \label{P101_eso_d2}
    \end{subfigure}
    \hfill
    \begin{subfigure}{0.32\textwidth} 
        \centering
        \includegraphics[width=\linewidth]{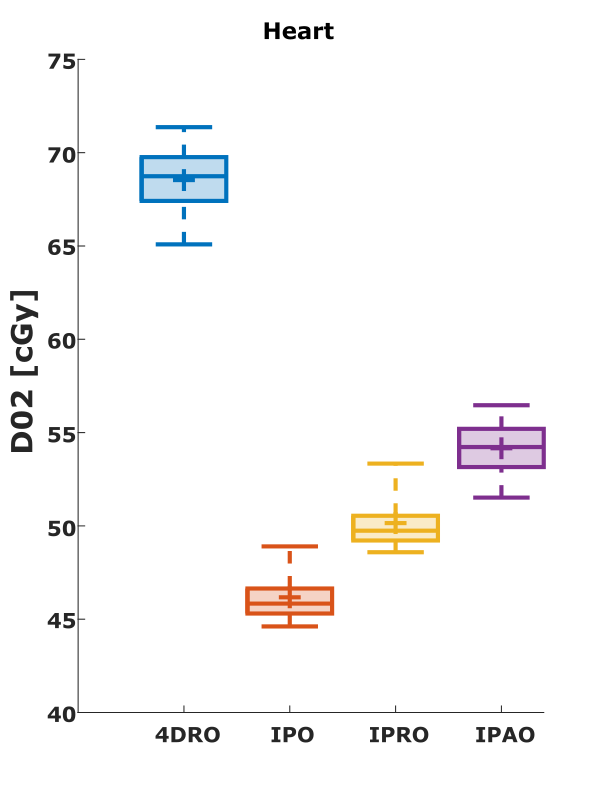}
        \caption{P111 - Heart}
        \label{P111_heart_d2}
    \end{subfigure}
    \begin{subfigure}{0.32\textwidth}
        \centering
        \includegraphics[width=\linewidth]{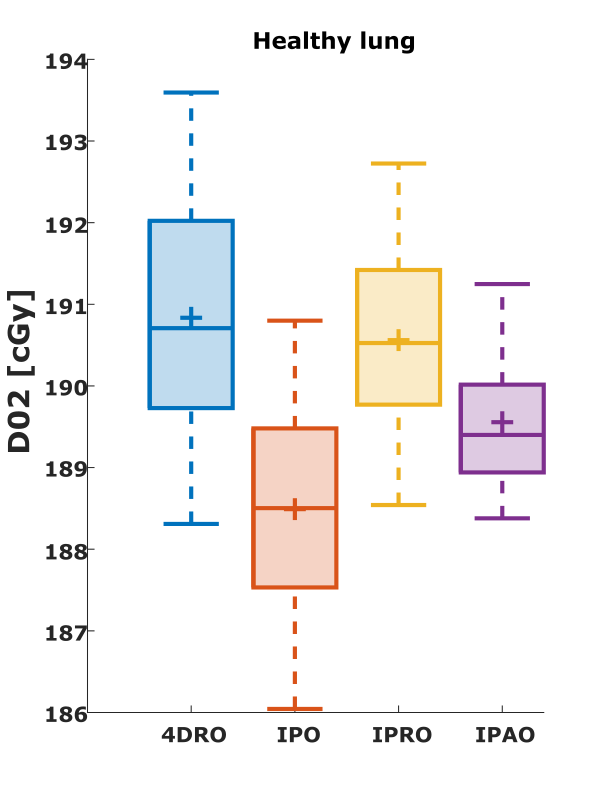}
        \caption{P111 - Healthy lung}
        \label{P111_lung_d2}
    \end{subfigure}
    \begin{subfigure}{0.32\textwidth}
        \centering
        \includegraphics[width=\linewidth]{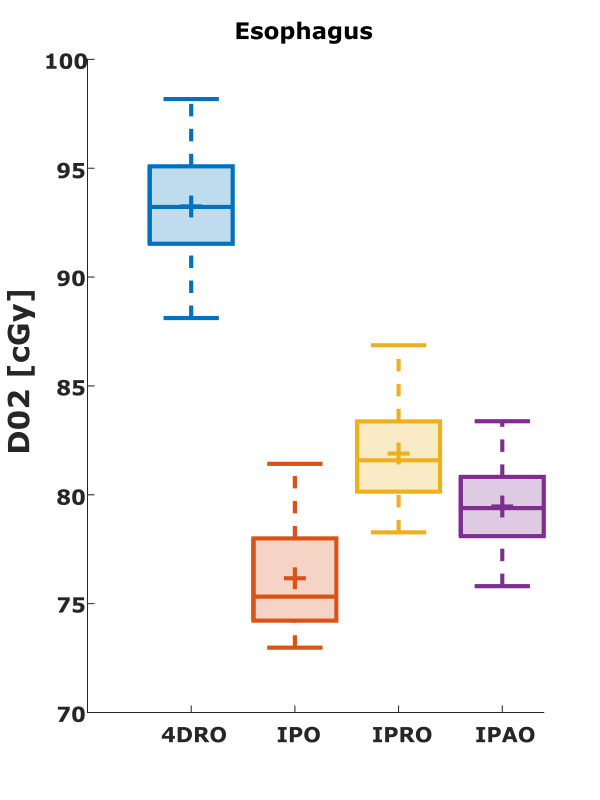}
        \caption{P111 - Esophagus}
        \label{P111_eso_d2}
    \end{subfigure}
    \hfill
    \begin{subfigure}{0.32\textwidth} 
        \centering
        \includegraphics[width=\linewidth]{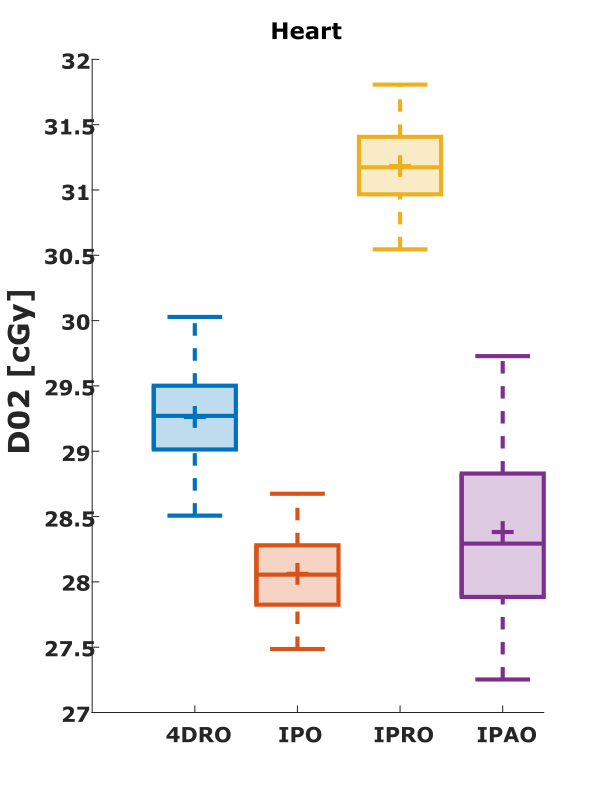}
        \caption{P114 - Heart}
        \label{P114_heart_d2}
    \end{subfigure}
    \begin{subfigure}{0.32\textwidth}
        \centering
        \includegraphics[width=\linewidth]{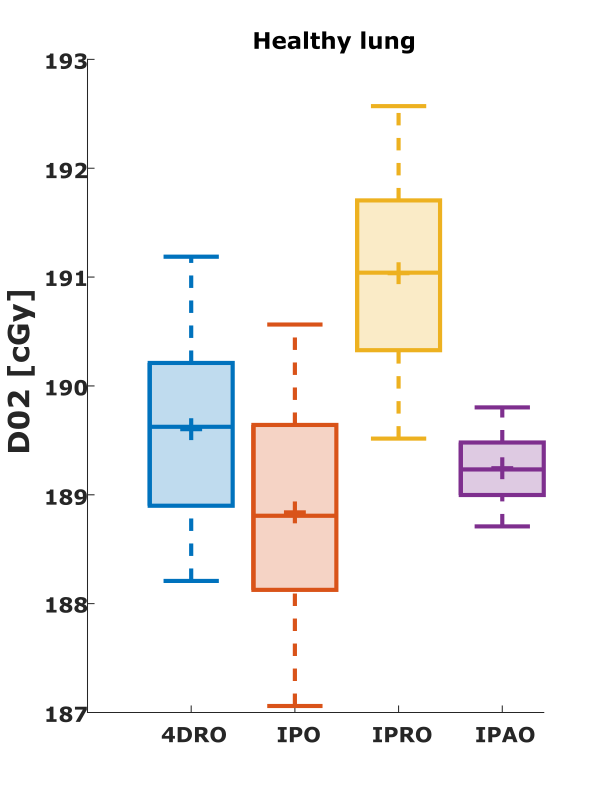}
        \caption{P114 - Healthy lung}
        \label{P114_lung_d2}
    \end{subfigure}
    \begin{subfigure}{0.32\textwidth}
        \centering
        \includegraphics[width=\linewidth]{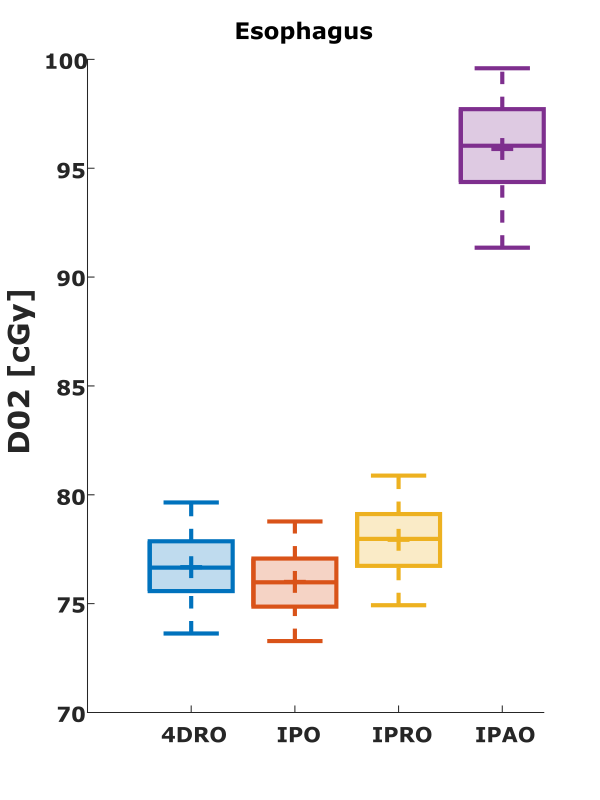}
        \caption{P114 - Esophagus}
        \label{P114_eso_d2}
    \end{subfigure}

    \caption{Box plots showcasing the spread of the OAR D2s for each patient. The box edges indicate quartiles, while the whiskers indicate 5th and 95th percentiles. The mean and median are indicated by the plus and the solid line, respectively.}
    \label{boxplots_oar_d2s}
\end{figure}

The methods are not as easily distinguished based on their doses to OARs. Instead, the best- and worst-performing methods vary between the patients, OARs, and metrics. The spreads of D2s for all combinations and OARs are shown in Figure \ref{boxplots_oar_d2s}. As a complement, the mean doses, for which the results were typically similar, are found in Appendix \ref{oar_doses}. In most cases, the lowest OAR doses were achieved with non-robust IPO. Among the remaining methods, the OAR results were the least distinguishable for P101; compared with 4DRO, IPRO and IPAO typically increased the dose to the heart and esophagus while marginally decreasing the dose to the healthy lung. These changes were, however, within 1.8 \% of the prescribed dose. For P111, the OAR doses were typically better for the interplay-driven methods than for 4DRO; the mean D2 decreased by 7.19--11.18 \% and by 5.68--8.64 \% of the prescription for the heart and esophagus, respectively (Figures \ref{P111_heart_d2} and \ref{P111_eso_d2}). For P114, IPRO exhibited the highest heart doses. The mean healthy lung dose was decreased slightly with IPAO compared to the other methods (Appendix \ref{oar_doses}). The esophagus D2 increased considerably (9.6 \% of the prescription when compared to 4DRO) with IPAO.

\section{Discussion}

In this work, we have proposed and highlighted the potential of an RHC-based framework for modifying the spot weights based on information acquired during the delivery of PBS. Comparing this adaptive approach with non-adaptive methods indicates that it results in more robust near-minimum and maximum doses to the CTV under varying patient motion. In particular, the substantially reduced variation of the CTV D98 indicates that it can mitigate the motion-induced uncertainty about the target coverage. As such, it stands out as an alternative to other computational motion management techniques such as 4DRO, IPRO, the plan library approach, and beam tracking.

An important conclusion from the results is that an adaptive re-optimization of PBS can be performed sequentially by including only a small subset of the remaining spots in each re-optimization. Here, we considered the spots in the subsequent energy layer as that subset. This design choice is intuitive for PBS, as the time window of the energy switch is a natural bottleneck in most systems and may be sufficiently long to allow modifications of the treatment plan. To investigate the impact of this reduction in degrees of freedom, we compared the results of IPO and IPAO in the nominal scenario. As we designed the motion prediction in IPAO to be completely accurate in the nominal scenario, these methods assume the same spot-to-phase assignment and the only difference is whether the energy layers are co-optimized (IPO) or optimized in sequence (IPAO). Interestingly, this difference did not severely affect the dosimetric performance, as revealed by their only marginally different nominal DVH curves (Figure \ref{dvh_bands}). This observation is important, as the limited horizon substantially reduces the computational cost of the optimization step and could be important to enable a real-time implementation.

Typically, OAR doses were not compromised by IPAO when improving the target coverage. Instead, the improvements in CTV D98 could be combined with improvements in CTV D2 while maintaining OAR doses at similar levels. Notably, all interplay-driven methods reduced the D2s of all the considered OARs for P111, compared to 4DRO. An exception was the esophagus dose for P114, which increased considerably for specifically IPAO compared to the other methods. Here, it is relevant to observe that P114 was also the patient for which the IPAO was the most dominant for both CTV D98 and CTV D2 (Figures \ref{P114_ctv_d98} and \ref{P114_ctv_d2}). This results suggests that although IPAO does not seem to increase OAR doses in general, there may be specific patient geometries for which a low-weighted OAR may increase in dose unexpectedly, probably when a higher-weighted objective, such as the target coverage, can be improved at its expense. This warrants caution when designing objective functions for real-time re-optimization purposes. However, it is reasonable to assume that results similar to those for the CTV would generalize and also hold for OARs, should they be weighted higher and treated robustly in the objective function.

For the non-adaptive interplay-driven optimization methods, the time structure of the delivery was updated every 10 iterations, as described in Algorithm \ref{alg2} and following the methodology in Bernatowicz et al.\ and Engwall et al.\ \cite{bernatowicz_advanced_2017, engwall_4d_2018}. This is a heuristic approach to managing the non-linear dependency of the dose calculation on the spot weights. In our experiments, however, we did not observe a large dependency of the optimization objective value on the time-structure update based on the updated spot weights. In addition, IPAO achieves its dosimetric results despite assuming a fixed time structure during optimization. Although we believe that initializing the optimizations with the spot weights from 4DRO contributes favorably in this regard, it is also largely dependent on the model of the delivery time structure, e.g., the importance of the spot weights and the assumed constant energy switching times. An increased dependency of the time structure on the spot weights, or accounting for uncertainty about an influential parameter such as the energy switching time, is expected to challenge this result, yielding more challenging optimization problems and potential problems with robustness. A broader investigation of all relevant uncertainties in combination, including those associated with the beam delivery, is needed for a complete understanding of the robustness aspects of a potential clinical implementation of any interplay-driven optimization approach. Alternatively, delivery concerns could be addressed by the specification of a fixed time structure with sufficient time between spots to make the uncertainties negligible, at the expense of increased treatment time. 

It should be said that much work remains before real-time adaptive PBS delivery using spot weight re-optimization can be implemented in practice in a TCS. In this work, we have based our 4DDCs on a single pre-treatment 4DCT per patient. We have done so to highlight the impact of the IPAO. A natural next step of the work is to apply the framework in a setting that involves irregular motion that varies also in amplitude. This extension poses new challenges that resemble those in a clinical implementation, particularly in terms of accuracy and speed of 4D dose computations, which is an active field of study \cite{duetschler_motion_2023, duetschler_limitations_2022}. In addition, the implementation must also include a method to track and predict the motion during the optimization horizon \cite{krieger_liver-ultrasound-guided_2021, bertholet_real-time_2019}. Although this, too, is challenging, it is facilitated by the fact that the motion must only be predicted accurately during the optimization horizon. Regardless, the dosimetric results will depend on the accuracy with which the motion prediction can be made. Finally, there is also the issue of information transfer within the subsystems involved in the delivery, information processing, and control. With reliable solutions to these problems, we believe that the dosimetric results presented here would remain unaffected, given that sufficiently many spots are available to cover the potential target motion. Then, there are certainly also more dedicated optimization methods that could be developed to enable finding good solutions within the time requirements of a real-time implementation.

\section{Conclusion}

From a dosimetric perspective, real-time re-optimization of PBS plans is an attractive motion management strategy. Under our modeling assumptions, the target coverage achieved by interplay-optimized plans without uncertainty can be maintained for a wide range of realistically variable breathing patterns.

\appendix

\section{OAR doses} \label{oar_doses}

\begin{figure}
    \centering
    \begin{subfigure}{0.32\textwidth} 
        \centering
        \includegraphics[width=\linewidth]{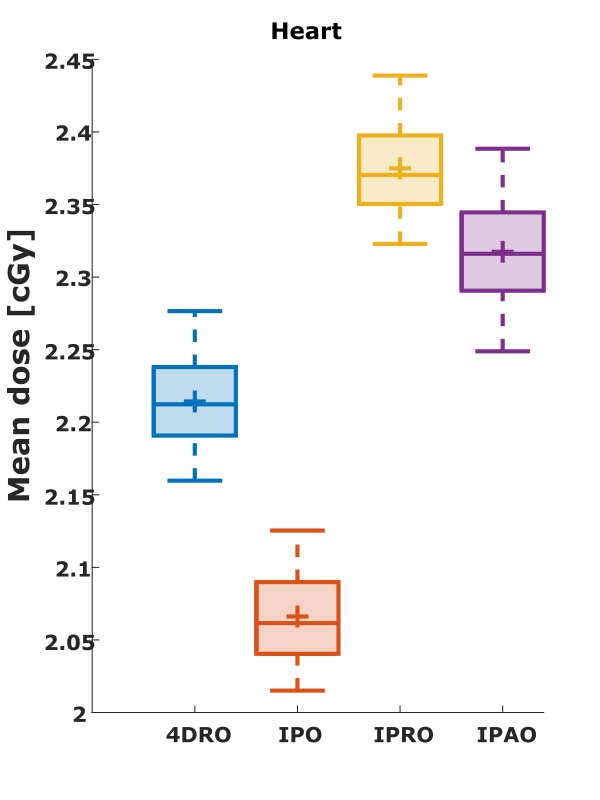}
        \caption{P101 - Heart}
        \label{P101_heart_mean}
    \end{subfigure}
    \begin{subfigure}{0.32\textwidth}
        \centering
        \includegraphics[width=\linewidth]{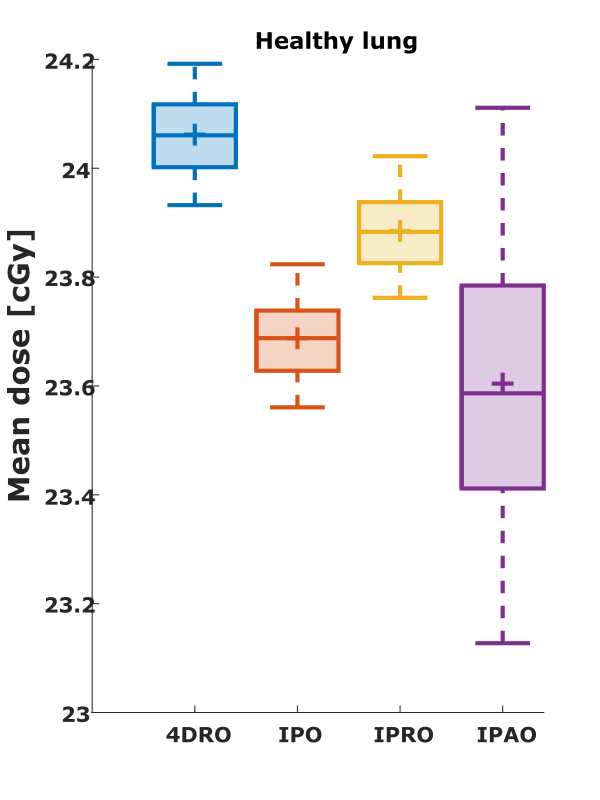}
        \caption{P101 - Healthy lung}
        \label{P101_lung_mean}
    \end{subfigure}
    \begin{subfigure}{0.32\textwidth}
        \centering
        \includegraphics[width=\linewidth]{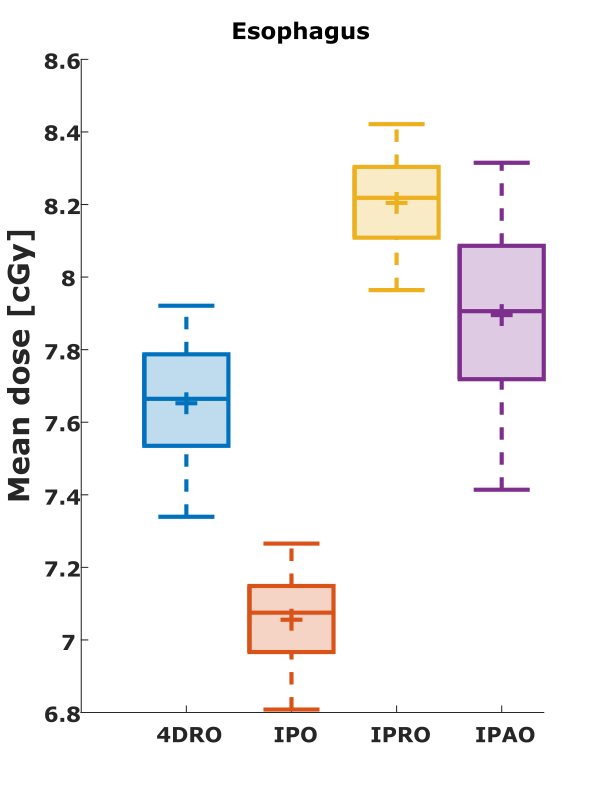}
        \caption{P101 - Esophagus}
        \label{P101_eso_mean}
    \end{subfigure}
    \hfill
    \begin{subfigure}{0.32\textwidth} 
        \centering
        \includegraphics[width=\linewidth]{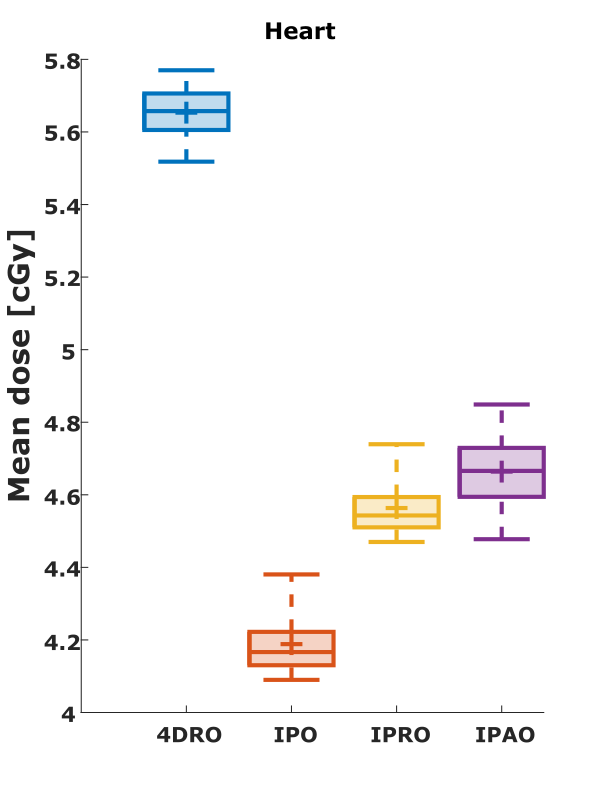}
        \caption{P111 - Heart}
        \label{P111_heart_mean}
    \end{subfigure}
    \begin{subfigure}{0.32\textwidth}
        \centering
        \includegraphics[width=\linewidth]{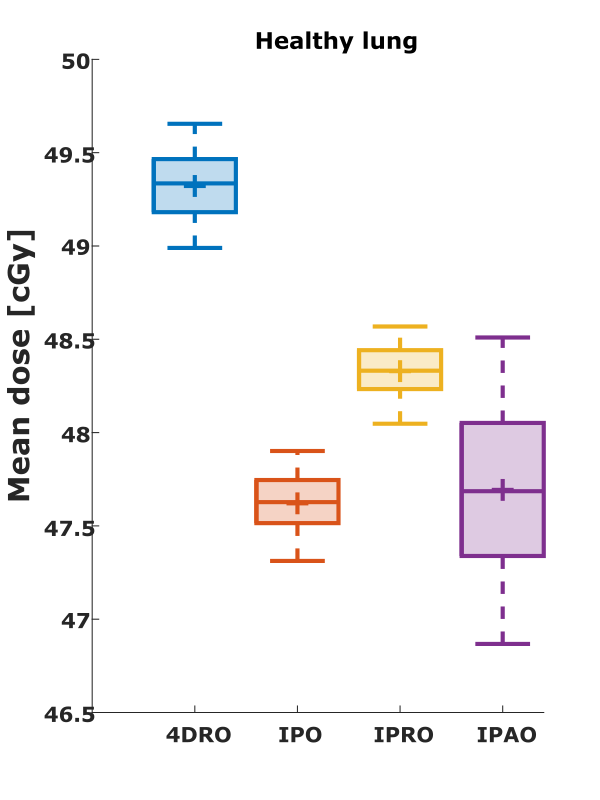}
        \caption{P111 - Healthy lung}
        \label{P111_lung_mean}
    \end{subfigure}
    \begin{subfigure}{0.32\textwidth}
        \centering
        \includegraphics[width=\linewidth]{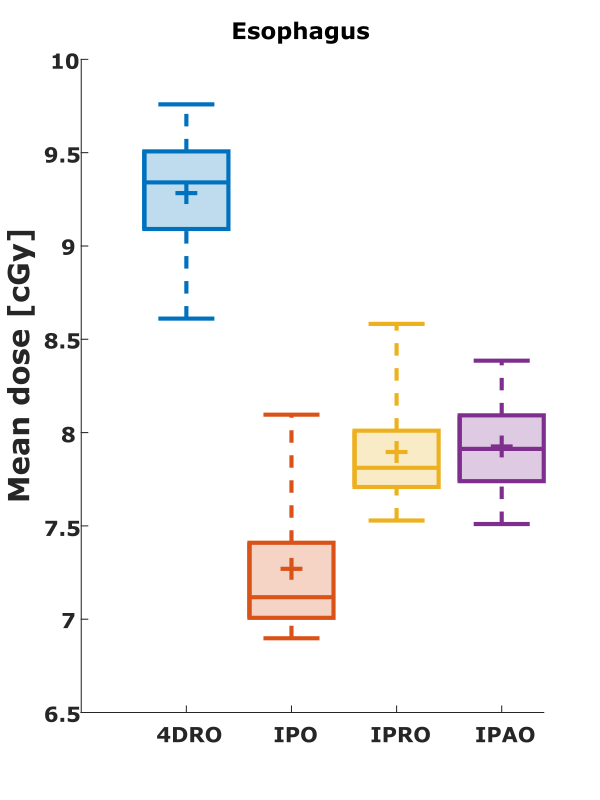}
        \caption{P111 - Esophagus}
        \label{P111_eso_mean}
    \end{subfigure}
    \hfill
    \begin{subfigure}{0.32\textwidth} 
        \centering
        \includegraphics[width=\linewidth]{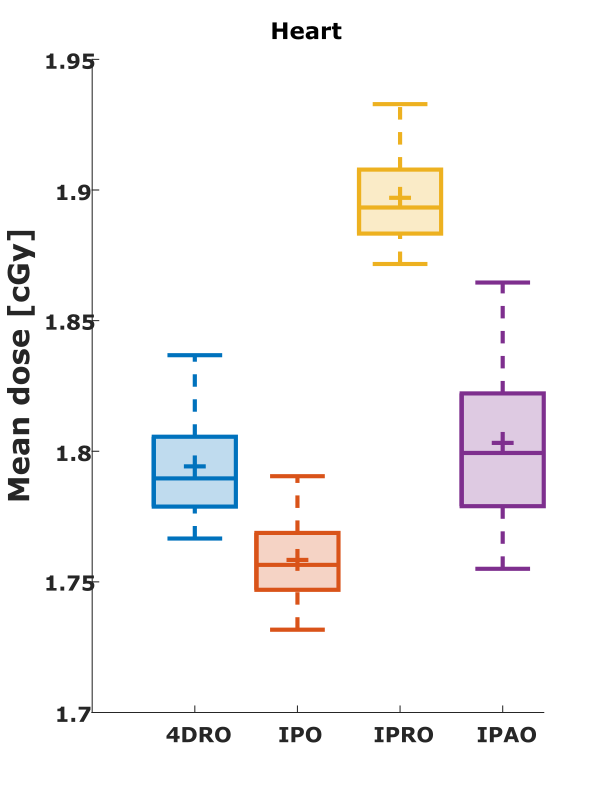}
        \caption{P114 - Heart}
        \label{P114_heart_mean}
    \end{subfigure}
    \begin{subfigure}{0.32\textwidth}
        \centering
        \includegraphics[width=\linewidth]{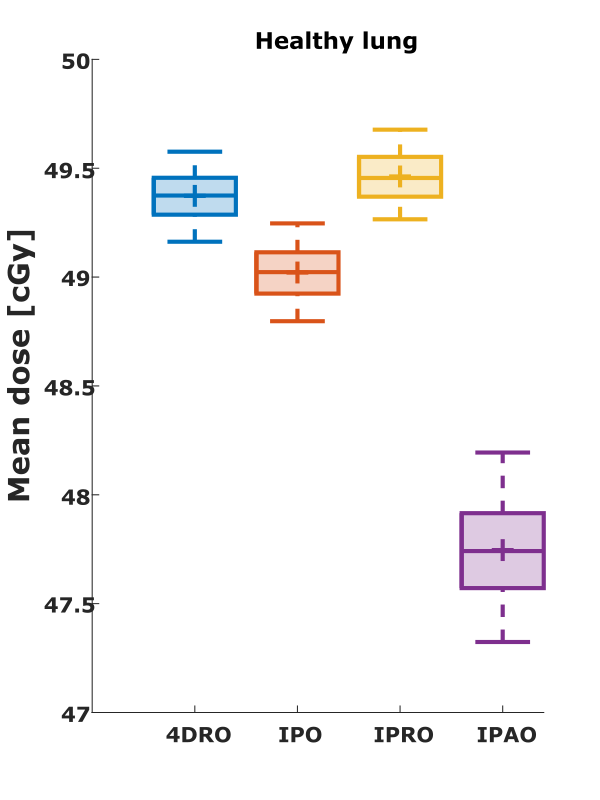}
        \caption{P114 - Healthy lung}
        \label{P114_lung_mean}
    \end{subfigure}
    \begin{subfigure}{0.32\textwidth}
        \centering
        \includegraphics[width=\linewidth]{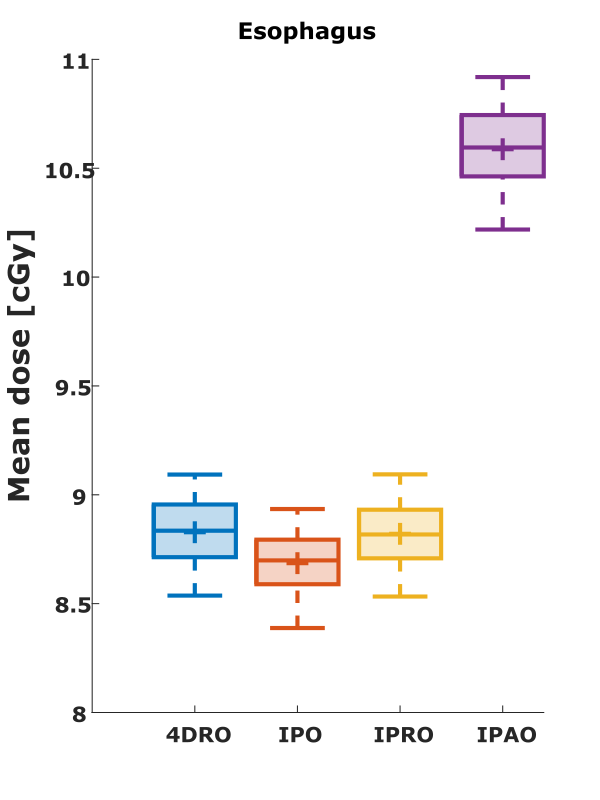}
        \caption{P114 - Esophagus}
        \label{P114_eso_mean}
    \end{subfigure}

    \caption{Box plots showcasing the spread of the OAR mean doses for each patient. The box edges indicate quartiles, while the whiskers indicate 5th and 95th percentiles. The mean and median are indicated by the plus and the solid line, respectively.}
    \label{boxplots_oar_mean}
\end{figure}

Figure \ref{boxplots_oar_mean} displays box plots of the mean dose to the OARs.

\printbibliography

\end{document}